\documentclass{article}

\usepackage{PRIMEarxiv}

\usepackage[utf8]{inputenc} 
\usepackage[T1]{fontenc}    
\usepackage{hyperref}       
\usepackage{url}            
\usepackage{booktabs}       
\usepackage{amsfonts}       
\usepackage{nicefrac}       
\usepackage{microtype}      
\usepackage{lipsum}
\usepackage{fancyhdr}       
\usepackage{graphicx}       
\graphicspath{{media/}}     

\title{
The Effectiveness of Graph Contrastive Learning on Mathematical Information Retrieval
}

\author{
  Pei-Syuan Wang, Hung-Hsuan Chen \\
  National Central University \\
  Taoyuan, Taiwan\\
  \texttt{peistu13333@gmail.com, hhchen1105@acm.org} \\
}

\usepackage[utf8]{inputenc} 
\usepackage[T1]{fontenc}    
\usepackage{hyperref}       
\usepackage{url}            
\usepackage{booktabs}       
\usepackage{amsfonts}       
\usepackage{nicefrac}       
\usepackage{microtype}      
\usepackage{xcolor}         
\usepackage{algorithm}
\usepackage[noend]{algpseudocode}
\usepackage{tikz}
\usetikzlibrary{shapes.multipart, positioning}
\usepackage{amsmath}
\usepackage{bm}
\usepackage{color}
\usepackage{graphicx}
\usepackage{makecell}
\usepackage{multirow}
\usepackage{subfigure}
\usepackage{tabularx}
\usepackage{mathtools}
\usepackage{pythonhighlight}
\usepackage{listings}
\usepackage{graphicx}
\usepackage{float}
\usepackage{subfigure} 
\usepackage{pdfpages}

\definecolor{pink}{rgb}{0.858, 0.188, 0.478}
\usepackage{xspace}

\newcommand{\figwidth}[4]{
    \begin{figure}[tb]\centering
    \includegraphics[width=#4]{fig/#1}
    \caption{#2}
    \label{#3}\end{figure}
    {}}

\newcommand{\figtwo}[9]{
    \begin{figure}[tb]\centering
    \subfigure[#2] {
    \label{#3}
    \includegraphics[width=#7in]{fig/#1}
    }
    \subfigure[#5] {
    \label{#6}
    \includegraphics[width=#7in]{fig/#4}
    }
    \caption{#8}
    \label{#9}
    \end{figure}
    {}}

\definecolor{commentcolor}{RGB}{110,154,155}   

\begin{document}
\maketitle

\begin{abstract}

This paper details an empirical investigation into using Graph Contrastive Learning (GCL) to generate mathematical equation representations, a critical aspect of Mathematical Information Retrieval (MIR). Our findings reveal that this simple approach consistently exceeds the performance of the current leading formula retrieval model, TangentCFT. To support ongoing research and development in this field, we have made our source code accessible to the public at \url{https://github.com/WangPeiSyuan/GCL-Formula-Retrieval/}.

\end{abstract}

\section{Introduction}

Search engines have revolutionized information access, enabling users to locate relevant textual content from the Internet quickly. Meanwhile, academic search engines and digital libraries, such as Google Scholar, CiteSeerX, and PubMed~\cite{caragea2014citeseer,wu2015citeseerx}, have become indispensable tools in the academic field, allowing researchers to discover related works from a vast amount of documents. Although a general-purpose search engine and an academic search engine may use different strategies to evaluate the quality of a document (e.g., a search engine may analyze the hyperlink structures to infer the importance of a webpage, while an academic search engine may rely on the citation counts to gauge the quality of a paper~\cite{chen2013csseer}), they often rely on similar strategies to define the relevance score between a query term and a document. Popular techniques include term frequency statistics (e.g., TFIDF and its variants~\cite{hsu2021toward}) and distributed representations learning (e.g., Word2Vec, fastText, and Transformer~\cite{mikolov2013distributed}).

Mathematical formulas commonly play a central role in scientific papers, facilitating the precise expression of abstract ideas. It is crucial to develop methodologies that can effectively retrieve documents that contain mathematical formulas similar to a target formula. Unfortunately, the search for mathematical formulas is very different from a regular text-based search. While textual search algorithms focus primarily on word frequencies, syntactic structures, and semantic associations, formula search requires a more profound comprehension of mathematical expressions, their inherent structures, and relationships between mathematical entities. Developing mathematical information retrieval (MIR) algorithms involves two main challenges. First, the model needs to capture the notation structure effectively. In MIR, the notation structure is perhaps more critical than string matching and term frequencies. For example, the quadratic equations $ax^2+bx+c = 0$ and $ \alpha \theta^2 + \beta \theta + \gamma = 0$ may convey the same concept, although they contain very different symbols. However, their structures are identical if we represent both equations using parse trees. Second, the labeled relevance score between pairs of mathematical formulas is needed for supervised training. Unfortunately, such datasets are limited. As a result, it could be difficult to apply machine-learned ranking (a.k.a. learning-to-rank)~\cite{liu2009learning} methodologies. These obstacles make MIR still an extremely challenging task.

In this paper, we experiment with applying graph contrastive learning (GCL) on the graph generated from the formula structure to capture the notation structure without the help of labeled relevance scores between formulas, thus addressing the above two challenges. We define the similarity score between each pair of formulas based on the cosine similarity between their embeddings. We conduct experiments using the NTCIR-12 MathIR Wikipedia Formula Browsing Task~\cite{zanibbi2016ntcir}. Experimental results show that our model consistently outperforms the TangentCFT model~\cite{mansouri2019tangent}, the state-of-the-art model for retrieving mathematical formulas. Note that our study focuses exclusively on models that rely solely on mathematical formulas for MIR. Therefore, models like MathBERT or Coco-MAE~\cite{peng2021mathbert,zhong2023mabowdor}, which also incorporate additional information such as contextual texts, fall outside the scope of our analysis.

\section{Related Work} \label{sec:rel-work}

This section reviews various methodologies in mathematics information retrieval, with a particular focus on TangentCFT, a state-of-the-art mathematics information retrieval method.

\subsection{Analyze formula using text}

Text-based MIR methods convert mathematical formulas into text formats such as \LaTeX and MathML and use text similarity measures to assess the similarity between formulas. An example of this approach is the TF-IDF-based method called Math Indexer and Searcher (MIaS)~\cite{sojka2011art}. It represents formulas in MathML format within an XHTML document and considers text and math formula components. However, this method largely overlooks mathematical formulas' structural and semantic aspects and relies mainly on a textual comparison based on words and their frequencies.

Other approaches employ complex natural language processing models to handle semantic retrieval. For example, Thanda et al.~\cite{thanda2016document} utilized the PV-DBOW model to learn the embeddings of text paragraphs. Gao et al.~\cite{gao2017preliminary} proposed the symbol2vec and formula2vec models, which are based on the Continuous Bag-of-Words (CBOW) and Doc2Vec architectures~\cite{mikolov2013distributed,le2014distributed}, respectively, to learn embeddings. These approaches generally transform formulas into vector representations using semantic representation methods. However, these methods minimally consider the structure of the formulas.

\subsection{Analyze formula using graph/tree}

Tree-based methods consider the symbol structure and arrangement of mathematical formulas by representing them in a structured format, such as a Symbol Layout Tree (SLT) or an Operator Tree (OPT), and compare the similarity of the structures to perform retrieval.

Among tree-based methods, some compared the similarity of two tree structures by matching paths from the root node to the child nodes~\cite{hijikata2007investigation,zhong2016opmes}. However, a successful match requires complete matching of root-to-leaf paths. Yokoi et al.~\cite{yokoi2009approach} propose a more flexible method by extracting subpaths from the root node to the child nodes and performing matching based on these subpaths, thus increasing the success rate of matching. The MCAT method~\cite{kristianto2016mcat} used both OPT and SLT for path extraction, incorporating path features and information about the sibling nodes, combined with text-based search, to achieve better results. Another method, Approach0~\cite{zhong2019structural}, used only OPT for path extraction, generating paths representing subexpressions of mathematical formulas. The similarity calculation is based on the largest common subexpression among the formulas.

\subsection{Integrating both formulas and contextual texts}

Some studies leveraged the formulas and contextual texts for math information retrieval. For example, MathBERT~\cite{peng2021mathbert}, motivated by the success of pre-trained language models in natural language processing, utilized math formula, its layout, and the contextual texts into a Transformer for training. Coco-MAE~\cite{zhong2023mabowdor} integrated the formula and textual information by contrastive learning. These studies leverage mathematical expressions and contextual texts, often leading to promising results in precision and recall.

\subsection{TangentCFT}

TangentCFT analyzes a formula based only on the formula but not the contextual text. TangentCFT begins by representing mathematical formulas using OPT and SLT. Next, TangentCFT traverses the tree and converts the paths in the tree into tuple sequences. These paths are then encoded and used to train embeddings using the fastText model~\cite{bojanowski2017enriching}. Finally, mathematical formula embeddings are obtained by averaging the embeddings of the tuples in a formula.

To the best of our knowledge, when considering the retrieval of mathematical formulas without leveraging contextual text information, TangentCFT is among the effective models currently available~\cite{peng2021mathbert}. Therefore, this paper uses TangentCFT as the baseline model and compares it with our approach.

\section{Methodology} \label{sec:method}

\figwidth{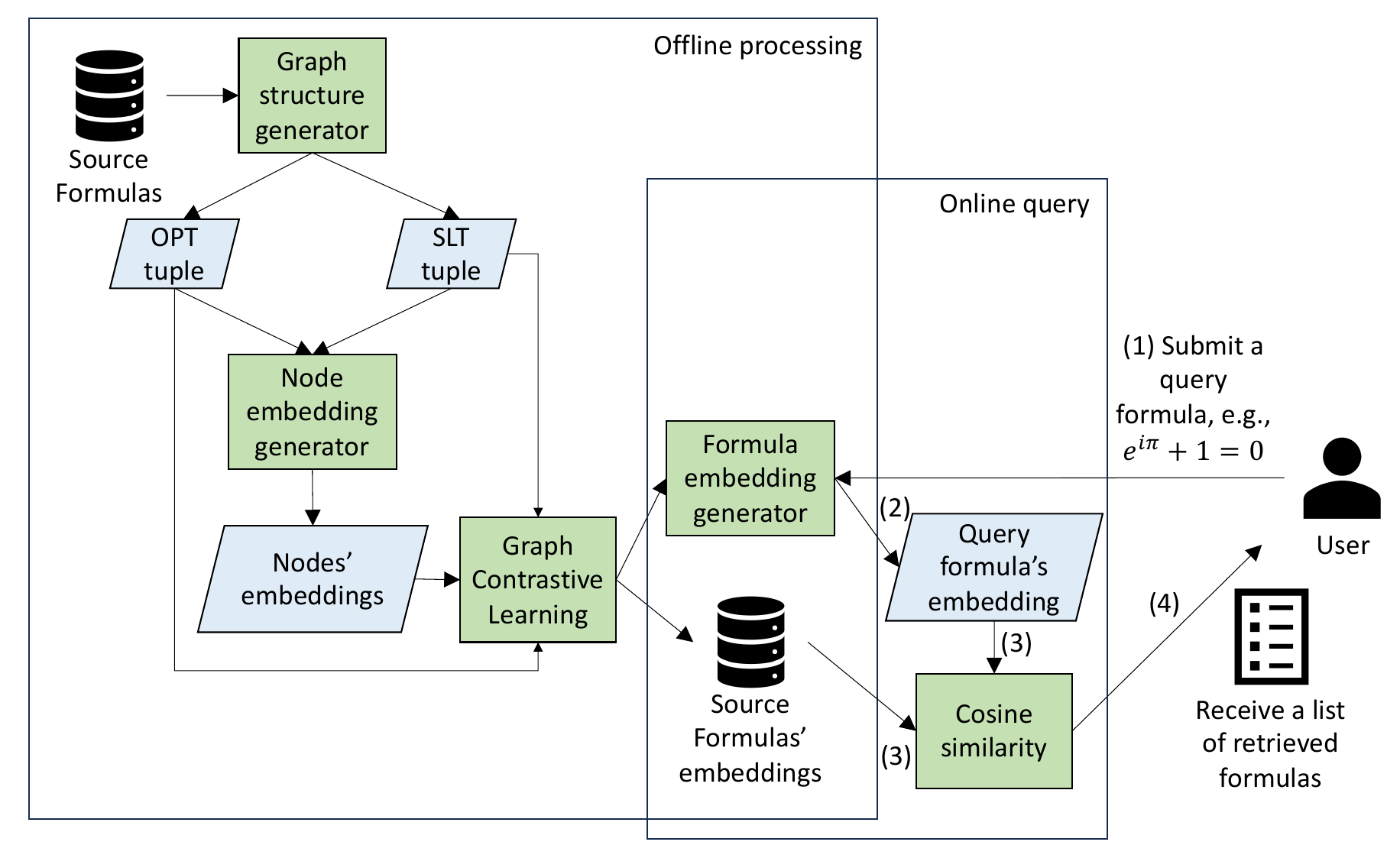}{The online and offline processing of the entire framework}{fig:flow}{0.9\textwidth}

We introduce the offline processing module and the online query module in this section. Figure~\ref{fig:flow} gives an overview of the whole workflow.

\subsection{Offline Processing Module}

The offline processing module includes a graph structure generator that outputs the OPT and SLT of a formula. We use TangentCFT to generate node (token) embedding, which will be the input of the graph contrastive learning models. The graph contrastive learning models generate formula embeddings based on contrastive learning; thus, the relevance scores between formula pairs are unnecessary. This section details the entire offline processing module.

\subsubsection{Graph Structure Generator}

A mathematical symbol sequence can form graphs expressing semantic relationships between symbols. This study employs two graph structures to represent the relationship of the symbols in a mathematical formula: Symbol Layout Tree (SLT) and Operator Tree (OPT)~\cite{davila2017layout}. The SLT is used primarily to indicate the spatial positioning of mathematical symbols in a written form. The OPT, on the other hand, is mainly used to capture the semantics of mathematical formulas. The OPT represents operators by an intermediate node, and the child nodes represent operands. Through the commutativity or associativity of operators, mathematically equivalent formulas with different appearances exhibit the same OPT structure.

\figtwo{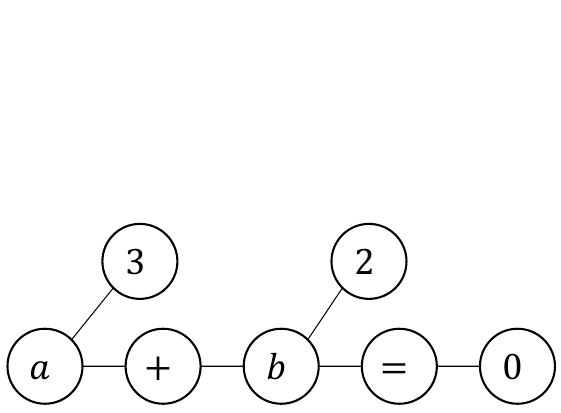}{SLT example}{fig:slt-exp}{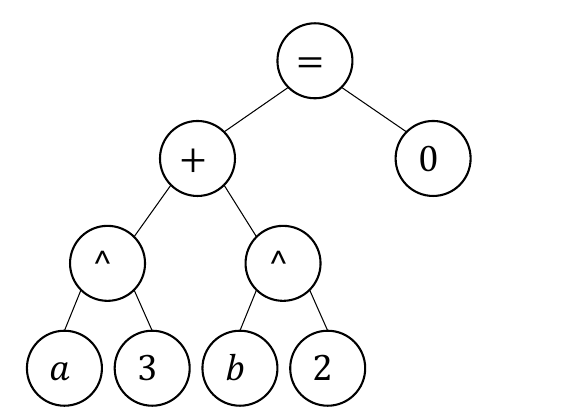}{OPT example}{fig:opt-exp}{1.4}{The examples of the SLT and OPT representations of the formula $a^3 + b^2 = 0$}{fig:slt-opt-exp}

For example, given a formula $a^3 + b^2 = 0$, Figure~\ref{fig:slt-opt-exp} gives its SLT and OPT representations: SLT generates a graph that better preserves the layout of the writing, whereas the output of OPT captures the semantics of the equation.

\subsubsection{Token Embedding Generator}

We may define the features for the nodes and edges of the SLT and OPT. For example, we could define a feature for a node that specifies whether the token represents an operator or operand. However, manually defining features can be tedious and perhaps subjective. Eventually, we decided to take advantage of the node and edge characteristics described in TangentCFT~\cite{mansouri2019tangent} and apply fastText~\cite{bojanowski2017enriching} to the paths sampled by random walks to generate the embeddings for the nodes. We set each output embedding length to $100$. These node embeddings are the building blocks for graph embeddings, which are representations of the formulas, as described below.

\subsubsection{Formula Embedding Generator and Graph Contrastive Learning}

\begin{table}[tb]
\caption{A comparison of the properties of the graph contrastive learning models.}
\label{tab:gcl-cmp}
\centering
\begin{tabular}{@{}cccc@{}}
\toprule
 & InfoGraph     & GraphCL        & BGRL         \\ \midrule
Requires negative pairs     & Y             & Y              & N            \\
Requires graph augmentation & N             & Y              & Y            \\
Contrastive pairs           & Graph to node & Graph to graph & Node to node \\
Number of encoders          & 1             & 1              & 2            \\ \bottomrule
\end{tabular}
\end{table}

To generate formula embeddings, we need to assemble the node embeddings. There are at least two different ways to do it. The first is to compute the elementwise average for each node embedding in a graph, as TangentCFT does~\cite{mansouri2019tangent}. However, the simple average may be too na\"{i}ve because the relationship among the math symbols (i.e., nodes) is missing. Another possibility is using the downstream task label as the ground truth and applying backpropagation to learn how to integrate the token embeddings. Unfortunately, our task has a limited number of relevance scores between pairs of formulas. Therefore, learning to integrate token embeddings based on a few labels will likely overfit the training data.

Eventually, we decided to employ graph contrastive learning methods to learn the embeddings of the formulas. GCL generates positive and negative graph pairs by manipulating the graph structures. Thus, the relevance score is not needed during training. We experimented with three representative GCL models: InfoGraph~\cite{sun2019infograph}, GraphCL~\cite{you2020graph}, and Bootstrapped Graph Latents (BGRL)~\cite{thakoor2021large}.  Since these models require no training labels, we can generate the formula embedding even if a formula does not appear in the training data.

The InfoGraph model processes multiple graphs in one batch. InfoGraph learns to generate the local node embeddings and the global graph embeddings simultaneously such that a node $n_i$ and a graph $g_j$ have high mutual information if $n_i \in g_j$ and low mutual information otherwise. An advantage of InfoGraph is that it does not rely on graph augmentation techniques. However, InfoGraph assumes that a node's embedding alone can discriminate its belonging graph and other graphs, which could be an over-strong assumption.

The GraphCL model generates a positive graph pair by augmenting a given graph based on, for example, node dropping and edge perturbation. 
GraphCL regards a negative graph pair by the augmented graphs of two distinct graphs. The loss function encourages positive graph pairs to have similar embeddings and negative graph pairs to have dissimilar embeddings. Although such a technique works exceptionally well in computer vision~\cite{chen2020simple}, the data augmentation techniques used in graphs may merit further discussion. For example, in image classification, an image after standard augmenting procedures (e.g., rotating or resizing) would still likely be regarded as having the same label. However, standard graph-augmentation techniques make the graphs structurally different, especially when a graph is small. As a result, the performance of GraphCL may be substantially influenced by the graph-augmenting procedure.

Finally, the BGRL requires only positive pairs generated by graph augmentation. BGRL alleviates the need for negative pairs by applying two distinct encoders: one's parameters are learned via direct backpropagation, and the other's parameters are updated by an exponential moving average of the parameters in the first encoder. Although BGRL is highly scalable because it requires no negative pairs, BGRL still needs graph augmentation, which could still be an issue, as discussed above.

Table~\ref{tab:gcl-cmp} compares these popular GCL models. Since each has its strengths and weaknesses, we tested all of them as formula embedding generation methods.

\subsection{Online Query Module}

A user submits to the system a query formula, which is used by the online query module to generate the query embedding based on the formula embedding generator trained offline. The system computes the cosine similarity between the query formula's embedding and each source formula's embedding. Finally, the system returns a list of the matched formulas by ranking the cosine similarities in descending order.
\section{Experiments} \label{sec:exp}

\subsection{Experimental Dataset and Evaluation Metrics}


We used the NTCIR-12 MathIR Wikipedia Formula Browsing Task~\cite{zanibbi2016ntcir} as the data source for evaluation. Each relevance score is an integer between 0 and 4.  

We used binary preference (bpref) and normalized discounted cumulative gain (nDCG) to evaluate the relevance of the returned formulas.  

The bpref score evaluates a binary retrieval task (relevant/irrelevant) with incomplete information, i.e., the relevance scores of some documents can be unlabeled~\cite{buckley2004retrieval}. The bpref is a perfect evaluation score in our case because the relevance scores between most pairs of documents are unlabeled in our dataset (we have only $1,202$ labeled relevance scores). We consider a pair of documents relevant if their relevance score is 3 or 4; otherwise, they are irrelevant.

The definition of bpref is given in Equation~\ref{eq:bpref}.

\begin{equation} \label{eq:bpref}
    s_{\textrm{bpref}} = \frac{1}{R}\sum_r \left(1 - \frac{\left|n \textrm{ is ranked higher than }r\right|}{\min(R, N)}\right),
\end{equation}
where $R$ and $N$ represent the counts of relevant and irrelevant documents, respectively, with $r$ as a relevant and $n$ as an irrelevant document.

Although a binary judgment (relevant/irrelevant) is probably more straightforward for human evaluation~\cite{kekalainen2005binary}, it fails to capture a fine-grained assessment. Therefore, we also applied the nDCG evaluation metrics because it allows for a graded relevance score. Equation~\ref{eq:ndcg} shows the formula of the nDCG score, which is the DCG score normalized by the ideal DCG score.

\begin{equation} \label{eq:ndcg}
    s_{\textrm{nDCG}} = \frac{s_\textrm{DCG}}{s_\textrm{IDCG}},
\end{equation}
where $s_\textrm{IDCG}$ is the score of $s_{\textrm{DCG}}$ when the top-$K$ documents are perfectly ordered (i.e., they are ordered according to the relevance score in descending order). The DCG score is computed by Equation~\ref{eq:dcg}.

\begin{equation} \label{eq:dcg}
    s_{\textrm{DCG}} = \sum_{i=1}^K \frac{r_i}{\log_2 (i+1)},
\end{equation}
where $r_i$ denotes the $i$th document's relevance score in the list ($r_i \in \{0, 1, 2, 3, 4\}$), and $K$ is the count of returned documents ($K=1,000$ in this experiment).

The nDCG is valid only with all returned documents scored for relevance. We filter out unjudged formulas from the list. Given each query has at most 90 judged formulas, and our $K=1,000$ exceeds this, we utilize all judged documents.

In general, the nDCG score measures the effectiveness of a ranking algorithm by considering the relevance and position of items in a ranked list. It places a higher emphasis on the top positions. Additionally, nDCG accommodates graded relevance judgments, allowing for finer distinctions in the relevance of items. Meanwhile, the bpref allows unjudged documents in the list, and the binary judgment is likely more intuitive for most evaluators. Since the two metrics assess the quality of ranked lists from different perspectives, we use both for evaluations.

\subsection{Quantitative result}

\begin{table}[tb]
\setlength{\tabcolsep}{5pt}
\caption{The bpref scores of applying different models on SLT layout, OPT layout, and F1 score of the above two. TangentCFT is the baseline; InfoGraph, BGRL, and GraphCL are GCL models used by our approach.}
\label{tab:bpref}
\centering
\begin{tabular}{@{}c|ccc@{}}
\toprule
Model       & SLT & OPT & F1 \\  \midrule
TangentCFT & $0.680\pm0.0053$ & $0.660\pm0.0064$ & $0.670$ \\
InfoGraph & $0.691\pm0.0066$ & $0.685\pm0.0070$ & $0.688$\\
BGRL & $\boldsymbol{0.701}\pm0.0089$ & $0.683\pm0.0077$ & $0.692$\\
GraphCL & $0.685\pm0.0090$ & $\boldsymbol{0.703}\pm0.0072$ & $\boldsymbol{0.694}$ \\ \bottomrule
\end{tabular}
\end{table}

\begin{table}[tb]
\setlength{\tabcolsep}{5pt}
\caption{The nDCG scores of applying different models on SLT layout, OPT layout, and F1 score of the above two. TangentCFT is the baseline; InfoGraph, BGRL, and GraphCL are GCL models used by our approach.}
\label{tab:ndcg}
\centering
\begin{tabular}{@{}c|ccc@{}}
\toprule
Model       & SLT & OPT & F1 \\  \midrule
TangentCFT & $0.841 \pm 0.0032$ & $0.830 \pm 0.0041$ & $0.835$ \\
InfoGraph & $\boldsymbol{0.860} \pm 0.0036$ & $0.851 \pm 0.0063$ & $0.855$ \\
BGRL & $0.851 \pm 0.0075$ & $0.827 \pm 0.0078$ & $0.839$ \\
GraphCL & $0.855 \pm 0.0029$ & $\boldsymbol{0.864} \pm 0.0065$ & $\boldsymbol{0.859}$ \\ \bottomrule
\end{tabular}
\end{table}

Table~\ref{tab:bpref} and Table~\ref{tab:ndcg} present the quantitative evaluation results of the bpref and nDCG scores when applying different GCL models on either the SLT or OPT layouts. We repeat each experiment 5 times and report the mean and standard deviation in these tables. We also report the F1 score for the mean of SLT and the mean of OPT scores. Both the bpref and the nDCG metrics indicate that our self-supervised graph contrastive learning consistently achieves better retrieval performance than TangentCFT, and the results are very stable (since the standard deviations are close to 0). In particular, the bpref score implies that, on average, the genuinely relevant formulas retrieved by our model rank higher than irrelevant ones more often when compared to the state-of-the-art TangentCFT. The nDCG scores also indicate that our method is better at ranking the most relevant formulas near the top.

Interestingly, various GCLs are more effective with different layouts: GraphCL works better when OPT is used, while InfoGraph and BGRL are more successful when using SLT. We show the F1 score in the last columns of Table~\ref{tab:bpref} and Table~\ref{tab:ndcg} to show the average effectiveness of each model on different layouts.

\subsection{Case Study}

\begin{table}[tbh]
\setlength{\tabcolsep}{5pt}
\caption{The top returns of various models when querying ``$O(mn \log m)$'' (using SLT as the layout for graph construction.)}
\label{tab:slt-case1}
\centering
\begin{tabular}{@{}cccc@{}}
\toprule
Rank & InfoGraph & GraphCL & BGRL \\ \midrule
1    & $O(mn \log m)$ & $O(mn \log m)$ & $O(mn \log m)$ \\
2    & $O(VE \log V)$ & $O(n \log m)$ & $O(m \log n)$\\
3    & $O(nk \log k)$ & $O(m \log n)$ & $O(n \log m)$\\
4    & $O(KN \log N)$ & $O(mn)$ & $O(mn)$\\
5    & $O(m + \log n)$ & $O(m^n)$ & $O(m^n)$\\ \bottomrule
\end{tabular}
\end{table}

\begin{table}[tbh]
\setlength{\tabcolsep}{5pt}
\caption{The top returns of various models when querying ``$O(mn \log m)$'' (using OPT as the layout for graph construction.)}
\label{tab:opt-case1}
\centering
\begin{tabular}{@{}cccc@{}}
\toprule
Rank & InfoGraph & GraphCL & BGRL \\ \midrule
1    & $O(mn \log m)$ & $O(mn \log m)$ & $O(mn \log m)$ \\
2    & $O(n \log m)$ & $O(n \log m)$ & $O(mn)$ \\
3    & $O(m \log n)$ & $O(mn \log (mn))$ & $O(Mr)$ \\
4    & $O(n \log k)$ & $O(m^2n \log n)$ & $O(mnp)$ \\
5    & $O(mn)$ & $O(m \log n \log \log n)$ & $\Theta(mn)$ \\ \bottomrule
\end{tabular}
\end{table}



\begin{table}[tbh]
\setlength{\tabcolsep}{5pt}
\caption{The top returns of various models when querying ``$\begin{bmatrix} V_1\\  I_2 \end{bmatrix} = \begin{bmatrix} h_{11} & h_{12}\\  h_{21} & h_{22} \end{bmatrix} \begin{bmatrix} I_1\\  V_2\end{bmatrix}$'' (using SLT as the layout for graph construction.)}
\label{tab:slt-case3}
\centering
\begin{tabular}{@{}cccc@{}}
\toprule
Rank & InfoGraph & GraphCL & BGRL \\ \midrule
1    & $\begin{bmatrix}V_{1}\\ I_{2}\end{bmatrix}=\begin{bmatrix}h_{11}&h_{12}\\ h_{21}&h_{22}\end{bmatrix}\begin{bmatrix}I_{1}\\ V_{2}\end{bmatrix}$ & 
$\begin{bmatrix}V_{1}\\ I_{2}\end{bmatrix}=\begin{bmatrix}h_{11}&h_{12}\\ h_{21}&h_{22}\end{bmatrix}\begin{bmatrix}I_{1}\\ V_{2}\end{bmatrix}$ &  
$\begin{bmatrix}V_{1}\\ I_{2}\end{bmatrix}=\begin{bmatrix}h_{11}&h_{12}\\ h_{21}&h_{22}\end{bmatrix}\begin{bmatrix}I_{1}\\ V_{2}\end{bmatrix}$ \\
2    & $\begin{bmatrix}V_{1}\\ I_{1}\end{bmatrix}=\begin{bmatrix}A&B\\ C&D\end{bmatrix}\begin{bmatrix}V_{2}\\ -I_{2}\end{bmatrix}$ &
$\begin{bmatrix}h_{11}&h_{12}\\ h_{21}&h_{22}\end{bmatrix}$ & 
$\begin{bmatrix}I_{1}\\ V_{2}\end{bmatrix}=\begin{bmatrix}g_{11}&g_{12}\\ g_{21}&g_{22}\end{bmatrix}\begin{bmatrix}V_{1}\\ I_{2}\end{bmatrix}$ \\
3    & $\begin{bmatrix}I_{1}\\ V_{2}\end{bmatrix}=\begin{bmatrix}g_{11}&g_{12}\\ g_{21}&g_{22}\end{bmatrix}\begin{bmatrix}V_{1}\\ I_{2}\end{bmatrix}$&  
$s_{(3,2,2,1)}=\begin{vmatrix}h_{3}&h_{4}&h_{5}&h_{6}\\ h_{1}&h_{2}&h_{3}&h_{4}\\ 1&h_{1}&h_{2}&h_{3}\\ 0&0&1&h_{1}\end{vmatrix}.$ & 
$\begin{bmatrix}V_{1}\\ I_{1}\end{bmatrix}=\begin{bmatrix}A&B\\ C&D\end{bmatrix}\begin{bmatrix}V_{2}\\ -I_{2}\end{bmatrix}$ \\
4    & $\begin{bmatrix}V_{1}\\ V_{2}\end{bmatrix}=\begin{bmatrix}0&-r\\ r&0\end{bmatrix}\begin{bmatrix}I_{1}\\ I_{2}\end{bmatrix}$ & 
$\begin{bmatrix}\dfrac{1}{h_{11}}&\dfrac{-h_{12}}{h_{11}}\\ \dfrac{h_{21}}{h_{11}}&\dfrac{\Delta\mathbf{[h]}}{h_{11}}\end{bmatrix}$ & 
$\begin{bmatrix}V_{2}\\ I^{\prime}_{2}\end{bmatrix}=\begin{bmatrix}1&-R\\ -sC&1+sCR\end{bmatrix}\begin{bmatrix}V_{1}\\ I_{1}\end{bmatrix}$ \\
5    & $\begin{bmatrix}h_{11}&h_{12}\\ h_{21}&h_{22}\end{bmatrix}$ & 
$h_{11}=\left.\frac{V_{1}}{I_{1}}\right|_{V_{2}=0}$ & 
$\begin{bmatrix}h_{11}&h_{12}\\ h_{21}&h_{22}\end{bmatrix}$ \\ \bottomrule
\end{tabular}
\end{table}

\begin{table}[tbh]
\setlength{\tabcolsep}{5pt}
\caption{The top returns of various models when querying ``$\begin{bmatrix} V_1\\  I_2 \end{bmatrix} = \begin{bmatrix} h_{11} & h_{12}\\  h_{21} & h_{22} \end{bmatrix} \begin{bmatrix} I_1\\  V_2\end{bmatrix}$'' (using OPT as the layout for graph construction.)}
\label{tab:opt-case3}
\centering
\begin{tabular}{@{}cccc@{}}
\toprule
Rank & InfoGraph & GraphCL & BGRL \\ \midrule
1    & $\begin{bmatrix}V_{1}\\ I_{2}\end{bmatrix}=\begin{bmatrix}h_{11}&h_{12}\\ h_{21}&h_{22}\end{bmatrix}\begin{bmatrix}I_{1}\\ V_{2}\end{bmatrix}$ & 
$\begin{bmatrix}V_{1}\\ I_{2}\end{bmatrix}=\begin{bmatrix}h_{11}&h_{12}\\ h_{21}&h_{22}\end{bmatrix}\begin{bmatrix}I_{1}\\ V_{2}\end{bmatrix}$  & 
$\begin{bmatrix}V_{1}\\ I_{2}\end{bmatrix}=\begin{bmatrix}h_{11}&h_{12}\\ h_{21}&h_{22}\end{bmatrix}\begin{bmatrix}I_{1}\\ V_{2}\end{bmatrix}$ \\
2 & $\begin{bmatrix}I_{1}\\ V_{2}\end{bmatrix}=\begin{bmatrix}g_{11}&g_{12}\\ g_{21}&g_{22}\end{bmatrix}\begin{bmatrix}V_{1}\\ I_{2}\end{bmatrix}$ & 
${I_{1}\choose I_{2}}=\begin{pmatrix}Y_{11}&Y_{12}\\ Y_{21}&Y_{22}\end{pmatrix}{V_{1}\choose V_{2}}$ & 
$\begin{bmatrix}I_{1}\\ V_{2}\end{bmatrix}=\begin{bmatrix}g_{11}&g_{12}\\ g_{21}&g_{22}\end{bmatrix}\begin{bmatrix}V_{1}\\ I_{2}\end{bmatrix}$ \\
3    & $\begin{bmatrix}V_{1}\\ V_{2}\end{bmatrix}=\begin{bmatrix}z_{11}&z_{12}\\ z_{21}&z_{22}\end{bmatrix}\begin{bmatrix}I_{1}\\ I_{2}\end{bmatrix}$ & 
$\begin{bmatrix}\dfrac{1}{h_{11}}&\dfrac{-h_{12}}{h_{11}}\\ \dfrac{h_{21}}{h_{11}}&\dfrac{\Delta\mathbf{[h]}}{h_{11}}\end{bmatrix}$ & 
$\begin{bmatrix}I_{1}\\ I_{2}\end{bmatrix}=\begin{bmatrix}y_{11}&y_{12}\\ y_{21}&y_{22}\end{bmatrix}\begin{bmatrix}V_{1}\\ V_{2}\end{bmatrix}$ \\
4    & $\begin{pmatrix}a_{1}\\ b_{1}\end{pmatrix}=\begin{pmatrix}T_{11}&T_{12}\\ T_{21}&T_{22}\end{pmatrix}\begin{pmatrix}b_{2}\\ a_{2}\end{pmatrix}$ & 
$\begin{bmatrix}h_{11}&h_{12}\\ h_{21}&h_{22}\end{bmatrix}$& 
$\begin{bmatrix}K_{11}&K_{12}\\ K_{21}&K_{22}\end{bmatrix}\begin{bmatrix}x_{1}\\ x_{2}\end{bmatrix}=\begin{bmatrix}F_{1}\\ F_{2}\end{bmatrix}$ \\
5    & $\begin{bmatrix}K_{11}&K_{12}\\ K_{21}&K_{22}\end{bmatrix}\begin{bmatrix}x_{1}\\ x_{2}\end{bmatrix}=\begin{bmatrix}F_{1}\\ F_{2}\end{bmatrix}$ & 
$\begin{bmatrix}\dfrac{\Delta\mathbf{[h]}}{h_{22}}&\dfrac{h_{12}}{h_{22}}\\ \dfrac{-h_{21}}{h_{22}}&\dfrac{1}{h_{22}}\end{bmatrix}$ & 
$\begin{pmatrix}A_{1}&B_{1}\\ A_{2}&B_{2}\end{pmatrix}\begin{pmatrix}x\\ y\end{pmatrix}=\begin{pmatrix}C_{1}\\ C_{2}\end{pmatrix}.$ \\ \bottomrule
\end{tabular}
\end{table}

This section shows the top retrieved formulas for two highly distinct query formulas. The first query involves a big-O expression with a logarithmic operation. Big-O notation is common in analyzing algorithms' time complexity. The inclusion of the $\log$ operation introduces mathematical complexity. Retrieving relevant results for such queries evaluates a model's capacity to deal with logarithmic functions, multiplications, and the big-O notation.
The second query is an equation represented in matrix form. Equations involving matrices are prevalent in various scientific and engineering fields, including linear algebra, physics, and computer graphics. Retrieving relevant results for matrix equations is essential in applications like solving linear systems or optimizing operations on large datasets.


Tables~\ref{tab:slt-case1} and~\ref{tab:opt-case1} show the top-5 returns of the GCL models for the query with the big-O and logarithm expression. All the best-matched formulas are precisely the query formula. Additionally, all returns involve the big-O notation, except the 5th return of BGRL using OPT, which retrieves a highly relevant big-$\Theta$ notation. Also, some formulas with semantics identical to ``$O(mn \log m)$'' but using different symbols, such as $O(VE \log V)$ or $O(KN \log N)$, are retrieved, indicating that these models effectively handles polynomials, logarithm, and the big-O notation.


Table~\ref{tab:slt-case3} and Table~\ref{tab:opt-case3} show the top-5 formulas retrieved from the query with the matrix equation. We arrive at the same findings as in the previous case. First, all the top returns are the same as in the query formula. Moreover, the top-5 returns of the models using SLT and OPT as the layout for graph construction all contain matrices, except the 5th return of GraphCL using SLT. In addition, models can retrieve semantically similar formulas with different symbols.



\section{Discussion} \label{sec:disc}

In this study, we investigate graph contrastive learning for formula retrieval to address two challenges of mathematical information retrieval: the model needs to capture the notation structure and the lack of relevance score between formula pairs. We explore the potential of popular GCL methods, including InfoGraph, GraphCL, and BGRL. We investigate the OPT and SLT graph layouts and their influence on the retrieval results. We observe that the GCL models outperform TangentCFT, a state-of-the-art formula retrieval model. However, TangentCFT is still essential, as our GCL models utilize the node embeddings generated by TangentCFT as the input for these GCL models. We also use case studies to confirm that the methods can handle different formula queries.

To enrich the training instances and further enhance model robustness, future work could explore the generation of new equations as positive training pairs based on equation templates. For example, given a regular expression that generates polynomial equations, each pair of these generated formulas could be a potential positive pair. Another future work could be to improve the GCL data-augmentation process. Current strategies, randomly adding or removing nodes/edges from graphs generated from equations, may not always be ideal. Such adjustments may alter the semantics of the equation and introduce structural inconsistencies. Thus, we are also interested in developing more sophisticated graph-augmentation strategies that preserve the meaning and structure of the equation while increasing the diversity of training data.

\bibliographystyle{unsrt}  
\bibliography{ref}

\begin{thebibliography}{10}

\bibitem{caragea2014citeseer}
Cornelia Caragea, Jian Wu, Alina Ciobanu, Kyle Williams, Juan Fern{\'a}ndez-Ram{\'\i}rez, Hung-Hsuan Chen, Zhaohui Wu, and Lee Giles.
\newblock Citeseer x: A scholarly big dataset.
\newblock In {\em Advances in Information Retrieval: 36th European Conference on IR Research, ECIR 2014, Amsterdam, The Netherlands, April 13-16, 2014. Proceedings 36}, pages 311--322. Springer, 2014.

\bibitem{wu2015citeseerx}
Jian Wu, Kyle~Mark Williams, Hung-Hsuan Chen, Madian Khabsa, Cornelia Caragea, Suppawong Tuarob, Alexander~G Ororbia, Douglas Jordan, Prasenjit Mitra, and C~Lee Giles.
\newblock Citeseerx: Ai in a digital library search engine.
\newblock {\em AI Magazine}, 36(3):35--48, 2015.

\bibitem{chen2013csseer}
Hung-Hsuan Chen, Pucktada Treeratpituk, Prasenjit Mitra, and C~Lee Giles.
\newblock Csseer: an expert recommendation system based on citeseerx.
\newblock In {\em Proceedings of the 13th ACM/IEEE-CS joint conference on Digital libraries}, pages 381--382, 2013.

\bibitem{hsu2021toward}
Li-Yuan Hsu, Chia-Hao Kao, I-Sheng Jheng, and Hung-Hsuan Chen.
\newblock Toward building an academic search engine understanding the purposes of the matched sentences in an abstract.
\newblock {\em IEEE Access}, 9:109344--109354, 2021.

\bibitem{mikolov2013distributed}
Tomas Mikolov, Ilya Sutskever, Kai Chen, Greg~S Corrado, and Jeff Dean.
\newblock Distributed representations of words and phrases and their compositionality.
\newblock {\em Advances in neural information processing systems}, 26, 2013.

\bibitem{liu2009learning}
Tie-Yan Liu et~al.
\newblock Learning to rank for information retrieval.
\newblock {\em Foundations and Trends{\textregistered} in Information Retrieval}, 3(3):225--331, 2009.

\bibitem{zanibbi2016ntcir}
Richard Zanibbi, Akiko Aizawa, Michael Kohlhase, Iadh Ounis, Goran Topic, and Kenny Davila.
\newblock Ntcir-12 mathir task overview.
\newblock In {\em NTCIR}, 2016.

\bibitem{mansouri2019tangent}
Behrooz Mansouri, Shaurya Rohatgi, Douglas~W Oard, Jian Wu, C~Lee Giles, and Richard Zanibbi.
\newblock Tangent-cft: An embedding model for mathematical formulas.
\newblock In {\em Proceedings of the 2019 ACM SIGIR international conference on theory of information retrieval}, pages 11--18, 2019.

\bibitem{peng2021mathbert}
Shuai Peng, Ke~Yuan, Liangcai Gao, and Zhi Tang.
\newblock Mathbert: A pre-trained model for mathematical formula understanding.
\newblock {\em arXiv preprint arXiv:2105.00377}, 2021.

\bibitem{zhong2023mabowdor}
Wei Zhong, Sheng-Chieh Lin, Jheng-Hong Yang, and Jimmy Lin.
\newblock One blade for one purpose: Advancing math information retrieval using hybrid search.
\newblock In {\em Proceedings of the 46th International ACM SIGIR Conference on Research and Development in Information Retrieval}, 2023.

\bibitem{sojka2011art}
Petr Sojka and Martin L{\'\i}{\v{s}}ka.
\newblock The art of mathematics retrieval.
\newblock In {\em Proceedings of the 11th ACM symposium on Document engineering}, pages 57--60, 2011.

\bibitem{thanda2016document}
Abhinav Thanda, Ankit Agarwal, Kushal Singla, Aditya Prakash, and Abhishek Gupta.
\newblock A document retrieval system for math queries.
\newblock In {\em NTCIR}, 2016.

\bibitem{gao2017preliminary}
Liangcai Gao, Zhuoren Jiang, Yue Yin, Ke~Yuan, Zuoyu Yan, and Zhi Tang.
\newblock Preliminary exploration of formula embedding for mathematical information retrieval: can mathematical formulae be embedded like a natural language?
\newblock {\em arXiv preprint arXiv:1707.05154}, 2017.

\bibitem{le2014distributed}
Quoc Le and Tomas Mikolov.
\newblock Distributed representations of sentences and documents.
\newblock In {\em International conference on machine learning}, pages 1188--1196. PMLR, 2014.

\bibitem{hijikata2007investigation}
Yoshinori Hijikata, Hideki Hashimoto, and Shogo Nishida.
\newblock An investigation of index formats for the search of mathml objects.
\newblock In {\em 2007 IEEE/WIC/ACM International Conferences on Web Intelligence and Intelligent Agent Technology-Workshops}, pages 244--248. IEEE, 2007.

\bibitem{zhong2016opmes}
Wei Zhong and Hui Fang.
\newblock Opmes: A similarity search engine for mathematical content.
\newblock In {\em Advances in Information Retrieval: 38th European Conference on IR Research, ECIR 2016, Padua, Italy, March 20--23, 2016. Proceedings 38}, pages 849--852. Springer, 2016.

\bibitem{yokoi2009approach}
Keisuke Yokoi and Akiko Aizawa.
\newblock An approach to similarity search for mathematical expressions using mathml.
\newblock {\em Towards a Digital Mathematics Library. Grand Bend, Ontario, Canada, July 8-9th, 2009}, pages 27--35, 2009.

\bibitem{kristianto2016mcat}
Giovanni~Yoko Kristianto, Goran Topic, and Akiko Aizawa.
\newblock Mcat math retrieval system for ntcir-12 mathir task.
\newblock In {\em NTCIR}, 2016.

\bibitem{zhong2019structural}
Wei Zhong and Richard Zanibbi.
\newblock Structural similarity search for formulas using leaf-root paths in operator subtrees.
\newblock In {\em Advances in Information Retrieval: 41st European Conference on IR Research, ECIR 2019, Cologne, Germany, April 14--18, 2019, Proceedings, Part I 41}, pages 116--129. Springer, 2019.

\bibitem{bojanowski2017enriching}
Piotr Bojanowski, Edouard Grave, Armand Joulin, and Tomas Mikolov.
\newblock Enriching word vectors with subword information.
\newblock {\em Transactions of the association for computational linguistics}, 5:135--146, 2017.

\bibitem{davila2017layout}
Kenny Davila and Richard Zanibbi.
\newblock Layout and semantics: Combining representations for mathematical formula search.
\newblock In {\em Proceedings of the 40th International ACM SIGIR Conference on Research and Development in Information Retrieval}, pages 1165--1168, 2017.

\bibitem{sun2019infograph}
Fan-Yun Sun, Jordan Hoffmann, Vikas Verma, and Jian Tang.
\newblock Infograph: Unsupervised and semi-supervised graph-level representation learning via mutual information maximization.
\newblock {\em arXiv preprint arXiv:1908.01000}, 2019.

\bibitem{you2020graph}
Yuning You, Tianlong Chen, Yongduo Sui, Ting Chen, Zhangyang Wang, and Yang Shen.
\newblock Graph contrastive learning with augmentations.
\newblock {\em Advances in neural information processing systems}, 33:5812--5823, 2020.

\bibitem{thakoor2021large}
Shantanu Thakoor, Corentin Tallec, Mohammad~Gheshlaghi Azar, Mehdi Azabou, Eva~L Dyer, Remi Munos, Petar Veli{\v{c}}kovi{\'c}, and Michal Valko.
\newblock Large-scale representation learning on graphs via bootstrapping.
\newblock {\em arXiv preprint arXiv:2102.06514}, 2021.

\bibitem{chen2020simple}
Ting Chen, Simon Kornblith, Mohammad Norouzi, and Geoffrey Hinton.
\newblock A simple framework for contrastive learning of visual representations.
\newblock In {\em International conference on machine learning}, pages 1597--1607. PMLR, 2020.

\bibitem{buckley2004retrieval}
Chris Buckley and Ellen~M Voorhees.
\newblock Retrieval evaluation with incomplete information.
\newblock In {\em Proceedings of the 27th annual international ACM SIGIR conference on Research and development in information retrieval}, pages 25--32, 2004.

\bibitem{kekalainen2005binary}
Jaana Kek{\"a}l{\"a}inen.
\newblock Binary and graded relevance in ir evaluations—comparison of the effects on ranking of ir systems.
\newblock {\em Information processing \& management}, 41(5):1019--1033, 2005.

\end{thebibliography}

\end{document}